\documentclass{article}
\usepackage{spconf,amsmath,graphicx,citesort}
\usepackage{enumitem}

\title{Real-Time Implementation Aspects of Large Intelligent Surfaces\vspace{-6pt}}
\name{Harsh Tataria, Fredrik Tufvesson, and Ove Edfors\vspace{-6pt}}
\address{Department of Electrical and Information Technology, Lund University, Lund, Sweden\\e-mail: \{harsh.tataria, fredrik.tufvesson, ove.edfors\}@eit.lth.se\vspace{-13pt}
\thanks{This work is partially supported by ELLIIT -- an Excellence Center at 
Linköping-Lund University in Information Technology, and Ericsson AB.}}

\begin{document}
\maketitle
\ninept

\begin{abstract}
\vspace{-1pt}
With the potential to provide a clean break from massive multiple-input multiple-output, \emph{large intelligent surfaces (LISs)} have recently received a thrust of research interest. Various proposals have been made in the literature to define the exact functionality of LISs, ranging from fully \emph{active} to largely \emph{passive} solutions. Nevertheless, almost all studies in the literature investigate the fundamental spectral efficiency performance of these architectures. In stark contrast, this paper investigates the \emph{implementation aspects} of LISs. Using the \emph{fully active} LIS as the basis of our exposition, we first present a rigorous discussion on the relative merits and disadvantages of possible implementation architectures from a \emph{radio-frequency circuits} and \emph{real-time processing} viewpoints. We then show that a \emph{distributed} architecture based on a common module interfacing a smaller number of antennas can be scalable. To avoid severe losses with analog signal distribution, multiple common modules can be interconnected via a digital nearest-neighbor network. Furthermore, we show that with such a design, the maximum backplane throughput scales with the number of served \emph{user terminals}, instead of the number of antennas across the surface. The discussions in the paper can serve as a guideline toward the real-time design and development of LISs.\vspace{-1pt}
\end{abstract}


\vspace{-8pt}
\section{Introduction}
\label{introduction}
\vspace{-6pt}
Massive multiple-input multiple-output (MIMO) systems are now starting to appear in commercial fifth-generation wireless networks around the world \cite{DAHMAN1}. With tens-to-hundreds of service antennas at a cellular base station (BS), massive MIMO can achieve significant spatial multiplexing gains \cite{BJORNSSON1,SHAFI1,SHAFI2,MARZETTA1,LARSSON1}. In addition, linear digital beamforming techniques to serve multiple user terminals within the same time-frequency resource have resulted in close to optimal performance over real propagation channels \cite{GAO1}. Relative to fourth-generation systems, this has resulted in an order-of-magnitude increase in the system spectral efficiency. Since the inception of massive MIMO in 2010 \cite{MARZETTA1}, phenomenal progress has been made in order to understand its theoretical and  implementation aspects (see e.g., \cite{BJORNSSON1,SHAFI1,LARSSON1,MARZETTA1,GAO1,NGO1,VIERA1,BJORNSSON2} for a taxonomy). To this end, our understanding of massive MIMO systems has greatly matured over almost a decade.  
\begin{figure}[!t]
    \centering
    \vspace{8pt}
    \hspace{-10pt}
    \includegraphics[width=7.3cm]{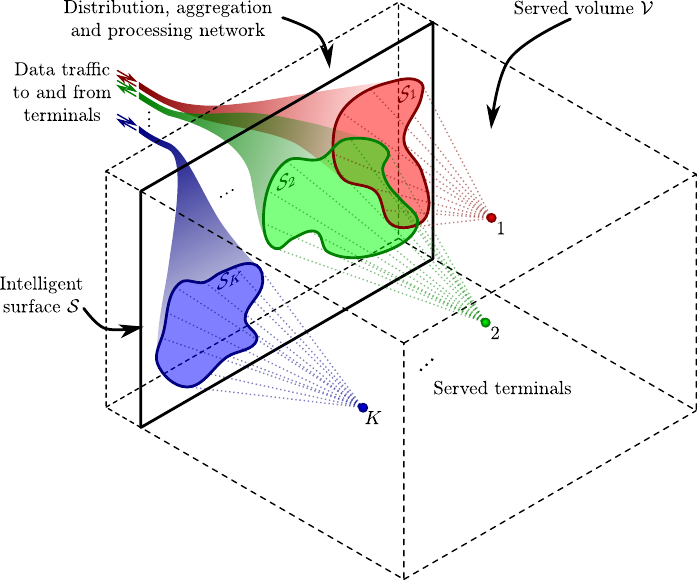}
    \vspace{-10pt}
    \caption{A fully active LIS, $\mathcal{S}$, serving $K$ randomly located terminals in volume $\mathcal{V}$ with the necessary backplane processing.}
    \label{LISfunctionality}
    \vspace{-18pt}
\end{figure}
\begin{figure*}[!t]
    \centering
    \vspace{-13pt}
    \includegraphics[width=13cm]{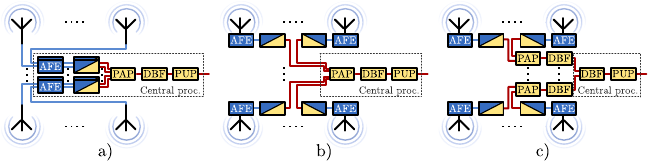}
    \vspace{-11pt}
    \caption{Three possible types of implementation architectures, where the \emph{analog} links are shown in \emph{blue} and the \emph{digital} links are shown in \emph{red} color. (a) Analog connections, (b) digital connections, and (c) digital connections with distributed beamforming functionality. The acronyms AFE, PAP, DBF and PUP denote analog front-end, (digital) per-antenna processing, (digital) beamforming, and (digital) per-user processing. The dual colored boxes (blue-yellow) denote AD and DA connections, respectively.}
    \label{LISArchitectures}
    \vspace{-14pt}
\end{figure*}

As a result, researchers in academia and industry have now began to investigate what lies \emph{beyond massive MIMO}. The authors in \cite{BJORNSSON1} highlight \emph{five} promising research directions for antenna array-based technologies. One possible direction is the use of a \emph{large intelligent surface (LIS)} \cite{HU1,LIAKOS1,SHI1}. As stated in \cite{HU1}, the original intention of LISs is that \emph{man-made structures} in the environment (such as building walls) which have substantially large physical apertures can be made \emph{electromagnetically active}. This implies that each part of the surface can transmit and receive electromagnetic radiation, acting as a BS with fully digital functionality. As a result of this, each antenna element on the surface needs to be able to transmit and receive radio-frequency (RF) signals. In this context, the term \emph{intelligent} implies that the surface is able to operate \emph{autonomously}, and is able to \emph{adapt} to changes in the propagation channel. Prior to the seminal work of \cite{HU1}, the authors of \cite{PUGLIELLI1} have made an attempt to study the impact of LISs on the overall communication system via University of California Berkley's \emph{e-Wallpaper} project. Since then, numerous studies have analyzed and evaluated the fundamental information-transfer capabilities of LISs, with and without hardware impairments, see e.g., \cite{HU1,HU2,SUBRT1}. Collectively, it has been shown that an \emph{infinite continuous} LIS can be replaced by its \emph{discrete} version \emph{without} loss in performance. Nonetheless, due to its large physical size and high backplane data throughput (generated from the many elements across the surface), such architectures may not be realized with \emph{centralized} processing. This is \emph{unlike} the case for massive MIMO systems where centralized processing, designed in a careful manner, does yield in practically implementable solutions \cite{VIERA1,SHEPARD1}. Henceforth, from a theoretical performance viewpoint,  research on \emph{distributed} architectures for LIS is under way \cite{HU3}, where a smaller number of distributed units, each having a separate signal processing chain are connected to a centralized processor. Figure~\ref{LISfunctionality} depicts the concept of an active LIS, $\mathcal{S}$, with the required backplane processing to serve $K$ terminals within a given three-dimensional volume $\mathcal{V}$. It is important note that the surface as a \emph{whole} does not need to serve a given terminal, since some terminals may be in closer proximity to a given part of the surface relative to others. Instead, for cost effective operation, smaller \emph{regions} (subsets) of the overall surface can be \emph{activated}, i.e., $\mathcal{S}_k\subseteq{}\mathcal{S}$ with $k=\left\{1,2,\dots,K\right\}$. Multiple activated areas do not need to be electrically connected, and different subsets may overlap, where the surface transmits and receives to/from more than one physical location at a given time. This helps to \emph{reduce} the per-link backplane data throughput, as well as to reduce the processing energy consumption - things which are extremely important when the surface aperture grows with a fixed number of served terminals.

The literature has proposed and analyzed the use of more \emph{passive intelligent reflecting surfaces (IRSs)}  \cite{WU1,WU2,WU3,HUANG1,HUANG2,DIRENZO1,MISHRA1,BJORNSSON3}. IRS is also often referred to as \emph{reconfigurable intelligent surfaces}, \emph{software controllable metasurfaces}, and \emph{reconfigurable reflect arrays}, respectively. The idea is to deploy an array of phase shifters (in hardware or software) across the IRS, which is located between a conventional BS and terminals for passive beamforming. At first glance, the passive operation seems appealing, since one may be able to manipulate the links between the BS and terminals without the use of RF up/down-conversion. However, \emph{synchronously} operating a large number of phase shifters with a low \emph{insertion loss} is a non-trivial task \cite{POZAR1}. Standard varactor-based or capacitive phase shifters are known to be rather lossy, and this loss scales at least linearly with the number of phase shifters \cite{UTTAM1}. More sophisticated phase shifters with adaptive tuning capability require \emph{active} control circuitry to inject a biasing voltage \emph{per-phase shifter}, greatly increasing the overall design complexity and energy consumption \cite{AN1}. Even more critically, in order to carry out passive beamforming from the IRS, one would have to compute the necessary beamforming weights based on some knowledge of channel state information. However, without active RF circuitry, there is no mechanism to sense the RF signals at the IRS. This implies that channel estimation can not be done passively. Furthermore, the fundamental performance benefits of IRS relative to conventional relays and massive MIMO systems are questionable (see e.g., \cite{BJORNSSON3,BJORNSSON4}). It has recently been proven that an IRS-based system \emph{can not} outperform a conventional massive MIMO system in terms of higher achievable signal-to-noise ratio (SNR) at a given terminal \cite{BJORNSSON4}. \emph{Due to these uncertainties in the performance aspects, we confine the discussions of the paper to the active LISs.}

Despite the rapid progress in assessing the performance limits of LISs, a study which discusses its implementation aspects and challenges is missing. This paper aims to close this gap. We first present a detailed discussion on the different contending implementation architectures for LISs, and evaluate their relative trade-offs. In terms of scalability, a distributed architecture across a number of common modules, where each module is connected to a smaller number of antennas may be the most suitable. We show that the decentralized nature of implementation can greatly simplify the analog signal distribution, and the physical surface layout. Multiple common modules can be connected with digital links, which are used to show that the resulting backplane throughput grows slowly as a function of the terminals, and \emph{not} the number of antennas on the LIS.

\vspace{-6pt}
\section{Real-Time Implementation Architectures}
\label{RealTimeImplementationArchitectures}
\vspace{-5pt}
In steep contrast to the massive MIMO, LISs require an unconventionally large physical area/size. For canonical frequencies up to the C-band (around 3.5 GHz), a 1024 element uni-polarized, half-wavelength spaced LIS with a 128$\times$8 panel array can easily occupy several meters of physical area in the horizontal dimension. This presents the first implementation challenge in \emph{efficiently routing} RF signals between the \emph{data source/sink} and the \emph{antenna elements} on the surface. This further raises \emph{two} closely related implementation questions: (1) How should the antenna elements across the LIS be connected to the central processing unit? (2) Is the required backplane bandwidth achievable in practice to facilitate the necessary signalling? Fundamentally, any transceiver architecture requires partitioning its functionality into blocks which need to be \emph{electrically} close to the radiating elements, and blocks which need to be near the centralized processor. Since the functionality of an RF transceiver can be split into the analog front-end, analog-to-digital (AD)/digital-to-analog (DA) converters, and baseband processing, there are \emph{three} qualitatively different architectures from distributing different portions of the transceiver chain. Following the same order as the arguments presented in \cite{PUGLIELLI2}, these can be summarized as: \begin{enumerate}[label=\alph*)]
    \item \textbf{Analog connections.} Here the AD, DA converters and all baseband hardware is at the central processor, while analog signals are routed from-and-to the radiating elements. 
    \item \textbf{Digital connections.} Each antenna element has a complete analog front-end, AD, DA converters and directly communicates  \emph{digitized} samples of the received analog waveform(s) with the central processing unit. 
    \item \textbf{Digital connections with distributed beamforming.} Here any \emph{per-antenna} baseband processing can be performed \emph{locally} at the element. Consequently, beamforming can be \emph{distributed} into computations performed across the LIS. With such an architecture, the signals exchanged with the central processor correspond to samples of the \emph{terminal} data, rather than each element's waveform. 
\end{enumerate}

The aforementioned architectures are presented in Fig.~\ref{LISArchitectures}, where a varying degree of centralized processing is observed as we transition from analog to the digital distributed processing. The analog architecture poses a couple of critical challenges that limit its use only to a small number of elements on the surface. Firstly, the routing of analog signals induces a loss that \emph{exponentially increases} with distance. Analog routing scales rather poorly with growing array size (and hence surface aperture), since a large number of elements require a greater number of transmission lines covering longer electrical distances. Rather interestingly, this issue is not alleviated by using higher quality transmission lines, due to their unusual form factor and cost (see \cite{ROCHA1}). The overall impact of the above is the reduction of received SNR at a terminal and at the surface. Secondly, analog architectures are highly sensitive to \emph{distortion} and \emph{external interference}, which limits the \emph{beamforming} performance to/from the surface. To this end, in order to minimize the length of analog tracks, each antenna element needs to be in close proximity to its associated data converters. Naturally, doing this lends itself to an architecture which composes of a grid of identical modules, each comprising of a single RF transceiver, as well as AD/DA converters. An alternative could be distribution of coherent RF signals over fiber \cite{AHMAD1}. Despite this, the signal distribution constitutes as a major challenge. Instead, \emph{multiple} transceivers can be fused into a \emph{common module} equipped with \emph{several} RF chains, AD/DA converters and a \emph{single} set of support hardware. This concept has been initially explored for massive MIMO systems \cite{PUGLIELLI2}, and here we extend it to the case of LIS . Rather interestingly, this implies that the optimal architecture contains a mixture of analog and digital processing, striking the right balance between sharing support hardware and analog routing loss. As such, the overall architecture could be formed with a \emph{multiplicity} of common modules tiled together with high speed digital interconnects to the nearest neighbors. Additionally, each common module is equipped with the digital hardware to perform \emph{distributed beamforming}, which substantially improves the scalability of the surface backplane network.
\begin{figure}[!t]
    \centering
    \hspace{-5pt}
    \includegraphics[width=8.7cm]{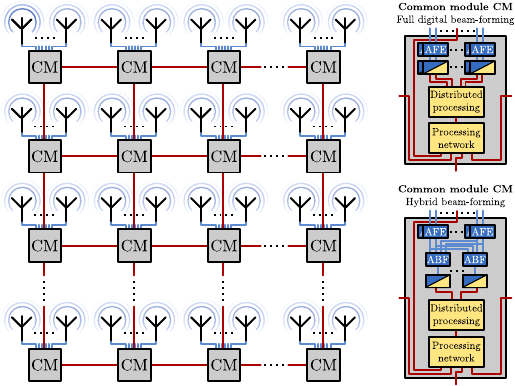}
    \vspace{-7pt}
    \caption{An example LIS architecture with a network of common modules. Analog links are depicted in \emph{blue} while the digital links are highlighted in \emph{red}. The acronyms are equivalent to Fig.~\ref{LISArchitectures}.}
    \label{DistributedBeamformingDesign}
    \vspace{-18pt}
\end{figure}
Figure~\ref{DistributedBeamformingDesign} depicts the block diagram of a LIS architecture consisting of common modules, as well as the two functional options for each common module in fully digital beamforming and hybrid (analog-digital) beamforming. 

In view of the above, the data throughputs generated or consumed by \emph{one antenna element} and \emph{one common module} on the LIS can be computed via the methodology in \cite{PUGLIELLI2}. To do this, we assume that the LIS contains an array of $M$ antennas serving $K$ terminals. The LIS can be divided into a discrete grid of $N$ modules, with $N$ being an integer divisor of $M$, such that each module drives $M/N$ radiating elements within a certain surface area. A system bandwidth $B$ is assumed such that the sampling rate of the AD/DA converters must be $\geq{}2B$. If each AD/DA converter has a resolution of $N_\textrm{res}$ bits, and is over-sampled by  $N_\textrm{over-samp}$, then the data throughput generated/consumed by a given element on the surface is $R=2BN_{\textrm{res}}N_{\textrm{over-samp}}$, and the throughput of one module is $\left(M/N\right)\hspace{-1pt}R$. The trade-off between the number of antennas and connected modules is an important one. This is since with a fixed $R$ and $N$, increasing $M$ yields an increase in the data throughput per module. To this end, an increased surface aperture with a higher number of antennas should be complimented with a relatively higher number of modules for efficient backplane processing.

\section{Backplane Interconnect Topologies and Impact of Distributed Beamforming}
\label{BackPlaneInterconnectTopologiesandDistributedBeamforming}
\begin{figure*}[!t]
    \centering
    \vspace{-11pt}
    \includegraphics[width=11cm]{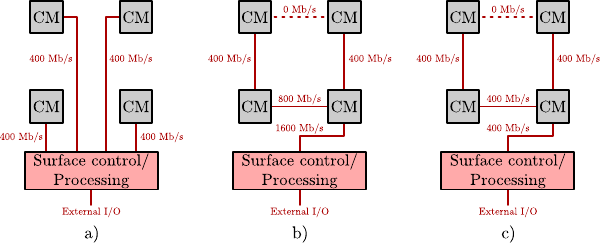}
    \vspace{-7pt}
    \caption{Possible routing schemes inspired by the arguments in \cite{PUGLIELLI2}. Throughputs are shown for an example where each block generates 400Mb/s. a) Fully parallel backplane interconnect. b) Mesh network with \emph{centralized} beamforming. c) Mesh network with \emph{distributed beamforming:} each block computes its estimate of the terminals' signals and these are summed throughout the routing network to generate the overall beamformed samples.}
    \label{RoutingSchemes}
    \vspace{-12pt}
\end{figure*}

\vspace{-5pt}
In a digitally interconnected architecture, the required backplane throughput serves the biggest hurdle limiting an increase in the number of antenna elements. As seen from the Lund University massive MIMO testbed, a total backplane throughput of 384 Gb/s was required to facilitate 100 elements over 20 MHz channel bandwidth \cite{VIERA1}. From this one can imagine the required throughput for a LIS containing hundreds-thousands of elements. We forsee \emph{two} types of interconnect topologies: First is when each common module can have a \emph{dedicated physical link} to the central control unit. Second is when all modules can be \emph{daisy-chained on a single link} over which all antenna signals are transmitted. These configurations seem to lean towards low per-link data throughput or low physical resource sharing. Naturally, other topologies can be designed by combining the two approaches, such as having multiple parallel chains. Moreover, a \emph{mesh network} can be considered as an extension of the daisy-chain concept, where each node can communicate with its nearest neighbors. The fully parallel backplane, as demonstrated in Fig.~\ref{RoutingSchemes} a) requires the lowest per-link data throughput. However, it has limitations in terms of scaliability. To serve all elements of the LIS, the interconnect length must grow with the size of the surface/array. This requires high performing, costly and high energy consuming links to support reliable transmission. In addition, routing complexity and the cross-interference between links also increases with the number of modules. The aforementioned challenges can be addressed by implementing the surface backplane as a nearest-neighbor mesh network, which require interconnects only at the scale of the \emph{inter-module distance}, regardless of the number of elements deployed on the surface.
Since the connections are only local, the challenge of globally routing $N$ links while maintaining acceptable cross-interference levels is avoided. The mesh network also presents a level of redundancy that allows for reconfiguration of the routing path to circumvent failures. 

Nonetheless, an interconnect topology consisting of a mesh network \emph{by itself} does not help to reduce the total required bandwidth at the surface control/processing level, as shown in Fig.~\ref{RoutingSchemes} b). When performing centralized beamforming, irrespective of the backplane topology, there is a fundamental requirement to exchange $M$ waveforms with the central processor, for a maximum data throughput of $M\hspace{-1pt}R$. Moreover, there is an additional penalty in the aggregate throughput due to the multi-hop nature of links. Assuming that the LIS contains $N$ modules which are connected to the central processor with $N_{\textrm{ch}}$ parallel daisy-chains, where $N_{\textrm{ch}}$ is an integer divisor of $N$. At any point along the chain, the data throughput is proportional to the number of \emph{preceding} elements. Thus, the \emph{aggregate} data throughput through the surface is given by \cite{PUGLIELLI2}
\vspace{-6pt}
\begin{equation}
    \label{DataOverallSurface}
    R_{\textrm{tot}}=N_\textrm{ch}\sum\limits_{\ell=1}^{\Delta}\ell
    \left(\hspace{-2pt}\frac{MR}{N}\hspace{-1pt}\right)
    \hspace{-1pt}=\hspace{-1pt}\frac{M}{2}
    \left(\Delta+1\right)R, 
    \vspace{-3pt}
\end{equation}
where $\Delta=N/N_\textrm{ch}$. In addition, the total power consumed by the backplane network \emph{increases} as the product of the number of antennas and the number of modules, corresponding to the penalty incurred by sending data through multiple hops. Similar effects take place in the fully parallel backplane since some links must communicate over a large distance, requiring increased energy consumption. To demonstrate the limitation of centralized routing, we consider a LIS with $M=1024$ elements with $B=20$ MHz, no oversampling, i.e., $N_\textrm{over-samp}=1$, and $N_{\textrm{res}}=10$. Under such modest conditions, neglecting additional system-related overheads, the throughput coming in-and-out of the central unit is 381.4 Gb/s. As stated in \cite{RAGHAVAN1}, a link operating at this rate over a few centimeters could achieve an energy efficiency of 1 pJ/bit. To this end, just transporting bits in and out of the central processor will consume hundreds of milliwatts of power. Even exploiting the greatest possible link parallelism at the central processor, it would be difficult to achieve LIS size-bandwidth product, especially for scalable architectures.  

The above problem can be resolved by distributing the processing at each module. This was originally suggested by \cite{SHEPARD2}, and was evaluated for distributed massive MIMO systems by \cite{PUGLIELLI2}. Below we extend these discussions for LIS, where we qualitatively compare the routing throughput, as well as discuss the impact of the common modules. The key observation is that the $M$ waveforms at the radiating elements of the LIS are not linearly independent but instead lie in a $K$-dimensional subspace generated by the $K$ distinctly located terminals \cite{PUGLIELLI2}. By exploiting this redundancy, it is possible to exchange only $K$ rather than $M$ unique signals with the central processor, performing distributed beamforming. Just as in massive MIMO systems, we forsee the regime of $M\gg{}K$, from which the required backplane throughput is substantially reduced. Since linear digital beamforming is performing matrix manipulations, such computations can be distributed. In the uplink, each element would multiply its received signal by a weight vector containing \emph{one entry per-terminal}. These vectors are then summed across the array to generate the per-terminal spatially filtered signals. This task can be embedded in the digital link to be low-latency and low-energy \cite{PUGLIELLI2}. The process is reversed in the downlink, where the terminal data streams are sent to all the modules and each element combines them with the appropriate beamforming weights to generate DA samples. 

\begin{figure}[!t]
    \centering
    \vspace{-10pt}
    \includegraphics[width=7.9cm]{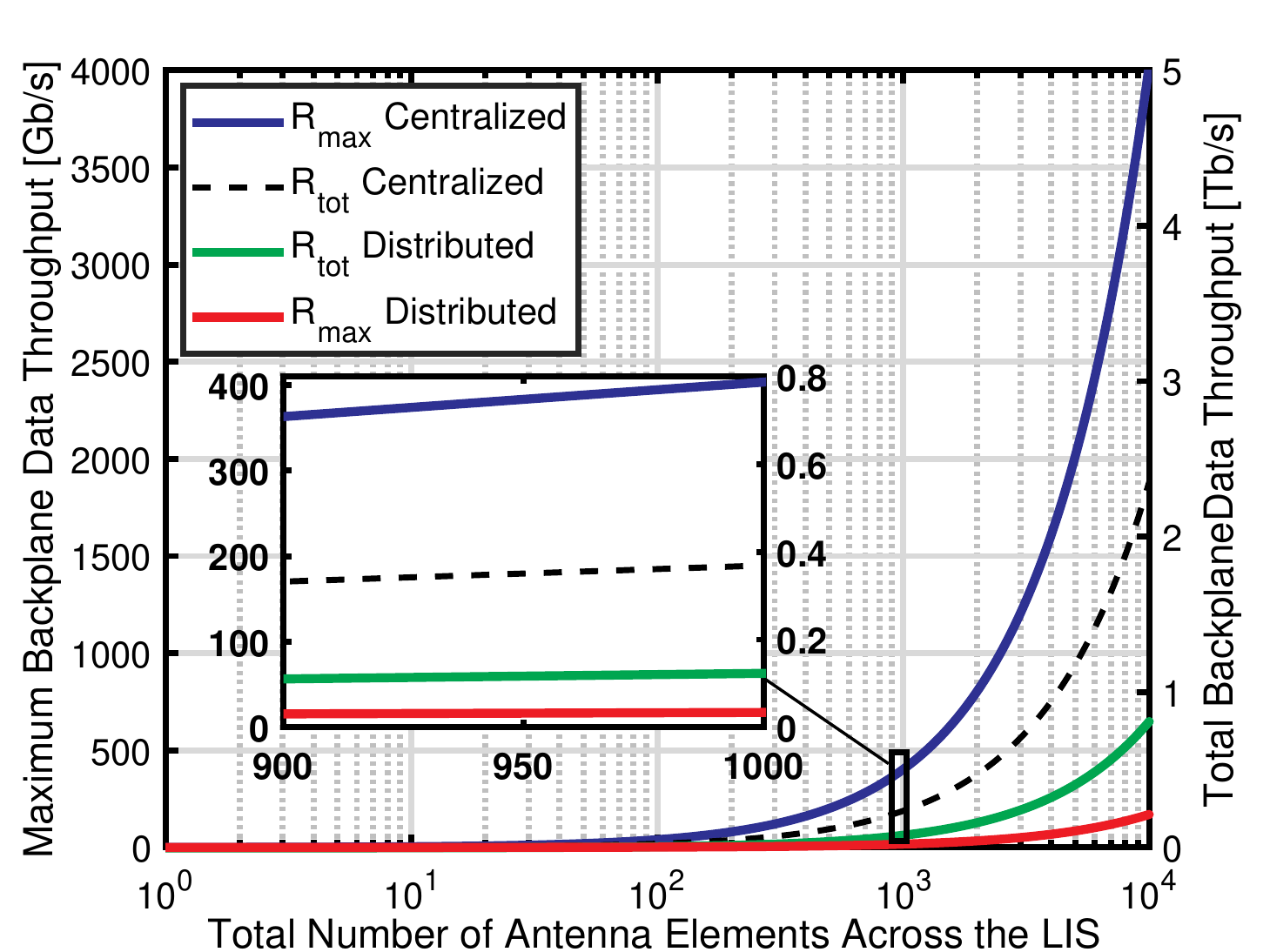}
    \vspace{-9pt}
    \caption{Maximum and aggregate backplane throughput with centralized and distributed beamforming for constant $M/K=32$.}
    \label{DataScaling}
    \vspace{-16pt}
\end{figure}

In an orthogonal frequency-division multiplexing-based system where beamforming is performed independently on each subcarrier, each common module would require a timing and frequency recovery block, downsampling and upsampling filters, a fast Fourier transform unit, as well as a small number of complex multipliers and adders. At the beginning of each frame, a given terminal transmits a synchronization and training preamble which can be used to estimate the timing parameters for each terminal at each module. This training preamble can also includes channel estimation for computation of beamforming weights. Subsequently, during data transmission, distributed linear beamforming is performed independently for each subcarrier. Note that in both uplink and downlink, the processing remains exactly the same. The only additional hardware required is a \emph{configurable-depth buffer} on each common module to match the latency of the backplane network \cite{PUGLIELLI2}. By routing the respective terminal  signals around the array rather than the antennas, \emph{the maximum required throughput is proportional to the number of terminals rather than the number of antennas on the LIS.} This is seen in Fig.~\ref{RoutingSchemes} c). We note that the above discussions are inspired by \cite{PUGLIELLI2}. We assume that each terminal's modulated data stream is represented by in-phase and quadrature samples of $N_\textrm{bit,beamf}$ bits each, then the maximum
throughput at the central processor is given by 
$R_{\textrm{max}}=2K\hspace{-1pt}BN_{\textrm{bit,beamf}}$,
and the aggregate throughput to deliver all $K$ signals to all $N$
modules is given by 
$R_{\textrm{tot}}=2N\hspace{-1pt}K\hspace{-1pt}BN_{\textrm{bit,beamf}}$ \cite{PUGLIELLI2}. The substantial improvement over fully centralized processing is evident since the throughput of the central processor's link is now proportional to the number of terminals. Figure~\ref{DataScaling} illustrates these benefits for an example with $B=20$ MHz, $N_{\textrm{res}}=10$, $N_{\textrm{bit,beamf}}$ = 15 and $N_{\textrm{ch}}=10$. With a constant ratio of $M/K =32$, both the maximum and aggregate throughputs are seen to reduce tremendously with distributed computations, highlighting its importance in such systems. 

\vspace{-20pt}
\section{Conclusion}
\label{Conclusion}
\vspace{-5pt}
The paper presents a first look at the implementation aspects of active LIS. An investigation into the relative merits and disadvantages of different architectures was carried out. To address the challenges associated with the large surface aperture and a large number of antennas, a distributed approach with a grid of common modules seems to serve as the most scalable solution. Each module is intended to drive a small number of elements and consists of a RF up/down-converter, data converters and a backplane connection to a central processor. The modules were digitally linked to the nearest neighbor via a mesh network, providing connectivity to the central processor. The maximum backplane communication throughput of a LIS was shown to slowly increase with the number of served terminals, providing an enormous advantage against centralized architectures.

\bibliographystyle{IEEEbib}

\end{document}